\documentstyle[12pt,psfig]{article}
\def\appendix{\par
  \setcounter{section}{0}
  \setcounter{subsection}{0}
  \def\@chapapp{\appendixname}%
  \def\thesection{\Alph{section}}}
\begin{document}
\author{A.N.Berlizov, A.A.Zhmudsky}
\date{\today}
\title{The recursive adaptive quadrature in MS Fortran-77}
\maketitle
\begin{abstract}
It is shown that MS Fortran-77 compilers allow to construct recursive
subroutines. The recursive one-dimensional adaptive quadrature subroutine
is considered in particular. Despite its extremely short body (only eleven
executable statements) the subroutine proved to be very effective and
competitive. It was tested on various rather complex integrands. The
possibility of function calls number minimization by choosing the optimal
number of Gaussian abscissas is considered.

The proposed recursive procedure can be effectively applied for creating
more sophisticated quadrature codes (one- or multi-dimensional) and easily
incorporated into existing programs.

\end{abstract}
\section{Introduction}
\label{intro}
As it was shown in \cite{BZh,Zh} the application of recursion makes it
possible to create compact, explicit and effective integration programs. In
the mentioned papers the C++ version of such a routine is presented.
However, it is historically formed that a large number of science and
engineering Fortran-77 codes have been accumulated by now in the form of
applied libraries and packages. That is one of the reasons why Fortran-77
is still quite popular in the applied programming. From this standpoint it
seems to be very useful to use such an effective recursive integration
algorithm in Fortran-77. There exist at least two possibilities to realize
it. The first one is described in \cite{BZh} where the interface for
calling mentioned C++ recursive integration function from MS Fortran-77 is
presented. The second possibility consist in constructing the recursive
subroutine by means of Fortran-77 only. This is the particular subject of
the paper where the possibility and benefits of recursion strategy in MS
Fortran-77 is discussed.

\section{Recursion in MS Fortran-77}
\label{recur}
The direct transformation of the mentioned C++ code is not possible mainly
due to the formal inhibition of the recursion in Fortran-77. However,
Microsoft extensions of Fortran-77 (e.g. MS Fortran V.5.0, Fortran Power
Station) allow to make indirect recursive calls. It means that subprogram
can call itself through intermediate subprogram. If anybody doubts he can
immediately try:

\begin{tabbing}
\= \hspace{7mm} \=                     \\
\> \>     call rec(1.0)                \\
\> \>     end                          \\
\> \>     subroutine rec(hh)           \\
\> \>     integer i/0/                 \\
\> \>     i = i + 1                    \\
\> \>     h = 0.5*hh                   \\
\> \>     write(*,*) i, h              \\
\> \>     if (i.lt.3) call mediator(h) \\
\> \>     write(*,*) i, h              \\
\> \>     end                          \\
\> \>     subroutine mediator(h)       \\
\> \>     call rec(h)                  \\
\> \>     end
\end{tabbing}
and get the following results:
\vspace{-5mm}
\begin{tabbing}
\= \hspace{7mm} \= \hspace{27mm}     \\
\>          1  \>   5.000000E-01   \\
\>          2  \>   2.500000E-01   \\
\>          3  \>   1.250000E-01   \\
\>          3  \>   1.250000E-01   \\
\>          3  \>   1.250000E-01   \\
\>          3  \>   1.250000E-01   \\
\end{tabbing}

But this is not a true recursion because no mechanism is supplied for
restoring the values of the internal variables of the subroutine after its
returning from recursion. The last requirement can be fulfilled by the
forced storing of the internal variables into the program stack. The
AUTOMATIC description of variables provides such possibility in MS
Fortran-77. Taking this into account, the above example can be rewritten:

\begin{tabbing}
\= \hspace{7mm} \=                     \\
\> \>     call rec(1.0)                \\
\> \>     end                          \\
\> \>     subroutine rec(hh)           \\
\> \>     integer i/0/                 \\
\> \>     automatic h, i               \\
\> \>     i = i + 1                    \\
\> \>     h = 0.5*hh                   \\
\> \>     write(*,*) i, h              \\
\> \>     if (i.lt.3) call mediator(h) \\
\> \>     write(*,*) i, h              \\
\> \>     end                          \\
\> \>     subroutine mediator(h)       \\
\> \>     call rec(h)                  \\
\> \>     end
\end{tabbing}
that yields:
\vspace{-5mm}
\begin{tabbing}
\= \hspace{7mm} \= \hspace{27mm}     \\
\>          1  \>   5.000000E-01   \\
\>          2  \>   2.500000E-01   \\
\>          3  \>   1.250000E-01   \\
\>          3  \>   1.250000E-01   \\
\>          3  \>   2.500000E-01   \\
\>          3  \>   5.000000E-01   \\
\end{tabbing}

Here the values of {\it h} are restored after each returning from recursion
because it is saved in the stack before the recursive call. Note, that
although the {\it i} variable is described as AUTOMATIC nonetheless its
value is not saved.

\section{Recursive adaptive quadrature algorithm}
The described possibilities allow to employ effective recursion strategy
for creating adaptive quadrature subroutine in MS Fortran-77.

\begin{figure}[h]
\begin{center}
\psfig{figure=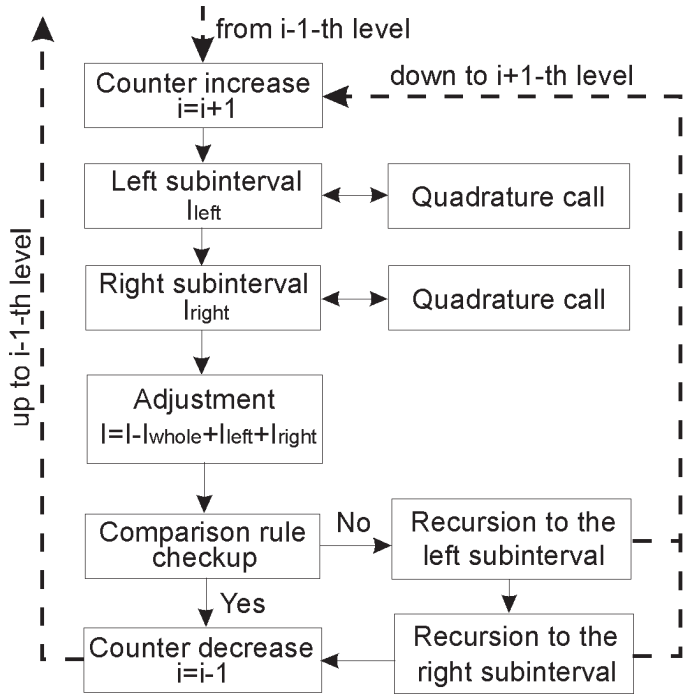,width=90mm}
\end{center}
\caption{\label{fig1}Adaptive recursive quadrature algorithm.}
\end{figure}

The presented algorithm consists of two independent parts: adaptive
subroutine and quadrature formula. The adaptive subroutine uses recursive
algorithm to implement standard bisection method (see fig.\ref{fig1}).
For reaching desired relative accuracy $\varepsilon$ of the integration the
integral estimation $I_{whole}$ over [$a_i$,$b_i$] subinterval on the i-th
step of bisection is compared with the sum of $I_{left}$ and $I_{right}$
integral values that are evaluated over left and right halves of the
considered subinterval. The comparison rule was chosen in the form:

\begin{equation} I_{left}+I_{right}-I_{whole}\leq \varepsilon\cdot I,
\label{pr:1}\label{qrule}\end{equation}
where $I$ denotes the integral sum over whole integration interval [a,b].
The value of $I$ is accumulated and adjusted on each step of bisection.

Should (\ref{qrule}) be not fulfilled the adaptive procedure is called
recursively for both (left and right) subintervals.
Evaluation of the integral sums on each step of bisection is performed by
means of quadrature formula. There are no restrictions on the type of
quadratures used during integration. This makes the code to be very
flexible and applicable to a wide range of integration problems.

The form (\ref{qrule}) of the chosen comparison rule does not pretend on
effectiveness rather on simplicity and generality. Really it seems to be
very common and does not depend on the integrand as well as quadrature
type. At the same time the use of (\ref{qrule}) in some cases can result in
overestimation of the calculated integral that consequently leads to more
integrand function calls. One certainly can get some gains using, for
instance, definite quadratures with different number or/and equidistant
points or Gauss-Kronrod quadrature \cite{Kron} etc. The comparison rule in
the later cases becomes more effective but complex, intricate and sometimes
less common. Whatever the case, the choice of comparison rule as well as
the problems connected with it lie outside the subject of the publication.

Let us note some advantages which the application of the recursive call
ideology to numerical integration can reveal:

\begin{itemize}
\item  Very simple and evident algorithm that could result in
extremely short as well as easy for further modifications and possible
enhancements adaptive code.
\item   Because of the indicated shortness the adaptive procedure's
own running time has to be diminutive. That could result in its better
performance compared to the known programs especially in the cases
when the integrand function calculations are not time consuming.
\item   There is no need to store the integrand function values and
pay attention on their correct usage. Besides no longer the control of
subinterval bounds is in need. Indicated features permit utmost
reduction of the efforts that one has to pay while creating the
adaptive code.
\item   Nothing but program's stack size sets the restriction on the
number of subintervals when the recursive procedure is used (see next
section). At the same time for the existing programs the crush level
of the primary interval is strictly limited by the dimensions of the
static arrays.
\end{itemize}

\section{Program realization}

Fortran-77 version of adaptive subroutine practically coincides with the
corresponding C++ QUADREC (Quadrature used
Adaptively and Recursively) function \cite{BZh}:
\begin{verbatim}
SUBROUTINE Quadrec(Fun,Left,Right,Estimation)
real*8 Fun, Left, Right, Estimation
real*8 Eps, Result, Section, SumLeft, SumRight, QuadRule
integer*4 RecMax, RecCur, RawInt
common /IP/ Eps, Result, RecMax, RawInt, RecCur
automatic SumLeft, SumRight, Section
external Fun
RecCur = RecCur+1
if (RecCur.le.RecMax) then
  Section = 0.5d0*(Left+Right)
  SumLeft = QuadRule(Fun,Left,Section)
  SumRight = QuadRule(Fun,Section,Right)
  Result = Result+SumLeft+SumRight-Estimation
  if (dabs(SumLeft+SumRight-Estimation).gt.Eps*dabs(Result)) then
    call Mediator(Fun,Left,Section,SumLeft)
    call Mediator(Fun,Section,Right,SumRight)
  end if
else
  RawInt = RawInt+1
end if
RecCur = RecCur-1
return
end
\end{verbatim}
Note that subroutine contains only eleven executable statements. The
integrand function name, left and right bounds of the integration interval
as well as the initial estimation of the integral value are the formal
arguments of the subroutine. The IP common block contains the following
variables: desired relative accuracy ({\it Eps}), the result of the
integration ({\it Result}), maximum and current levels of recursion ({\it
RecMax, RecCur}) as well as raw (not processed during integration)
subintervals ({\it RawInt}).

The {\it Section} variable is used for storing a value of midpoint of the
current subinterval. The integral sums over its left and right halves are
estimated by {\it QuadRule} external function and stored in {\it LeftSum}
and {\it RightSum} variables. The last three variables are declared as
AUTOMATIC allowing to preserve their values from changing and use them
after returning from recursion.

Execution of the subroutine begins with increasing of recursion level
counter. If its value does not exceed {\it RecMax}, the integral sums over
left and right halves of the current subinterval are evaluated, the
integration result is updated and accuracy of the integration is checked.
If achieved accuracy is not sufficient then subprogram calls itself (with
the help of mediator subroutine) over left and right halves of subinterval.
In the case the accuracy condition is satisfied the recursion counter
decreases and subprogram returns to the previous level of recursion. The
number of raw subintervals is increased when desired accuracy is not
reached and {\it RecCur} is equal to {\it RecMax}.

The mediator subroutine has only one executable statement:
\begin{verbatim}
      subroutine Mediator(Fun,Left,Right,Estimation)
      real*8 Estimation, Fun, Left, Right
      external Fun
      call Quadrec(Fun,Left,Right,Estimation)
      return
      end
\end{verbatim}

The main part of the integration program can be look like:
\begin{verbatim}
      common   /IP/ Eps, Result, RecMax, RawInt, RecCur
      common   /XW/ X, W, N
      integer  RecMax, RawInt, RecCur
      real*8   X(100), W(100), Left/0.0d0/, Right/1.0d0/
      real*8   Result, Eps, QuadRule
      real*8   Integrand
      external Integrand
      Eps = 1.0d-14
      N = 10
      RecCur = 0
      RecMax = 10
      call gauleg(-1.0d0,1.0d0,X,W,N)
      Result = QuadRule(Integrand,Left,Right)
      call Quadrec(Integrand,Left,Right,Result)
      write(*,*) ' Result = ',Result,'  RawInt = ',RawInt
      end
\end{verbatim}
The common block {\it XW} contains Gaussian abscissas and weights
which are calculated with the help of {\it gauleg} subroutine for a given
number of points {\it N}. The text of subroutine, reproduced from
\cite{Press}, is presented in Appendix \ref{GL}.

The text of {\it QuadRule} function is presented below:
\begin{verbatim}
      real*8 function QuadRule(Integrand,Left,Right)
      common /XW/ X, W, N
      real*8 X(100),W(100),IntSum,Abscissa,Left,Right,Integrand
      IntSum = 0.0d0
      do 1 i = 1, N
      Abscissa = 0.5d0*(Right+Left+(Right-Left)*X(i))
1     IntSum = IntSum + W(i)*Integrand(Abscissa)
      QuadRule = 0.5d0*IntSum*(Right-Left)
      return
      end
\end{verbatim}

It is important to note that the number of recursive calls is limited by
the size the program stack. This fact obviously sets the limit on the
reachable number of the primary integration interval bisections and
consequently restricts the integration accuracy. Note that stack size of
the program can be enlarged by using /Fxxxx option of MS Fortran-77
compiler.

\section{Numerical tests}

The program testing was performed on four different integrals. In each case
the exact values can be found analytically. That made it possible to
control the desired and reached accuracy of the integration. Besides the
same integrals were obtained with the help of well-known adaptive program
QUANC8 reproduced from \cite{Fors}. It allowed to compare the number of
integrand function calls and the number of raw intervals for both programs.

The presented comparison has merely the aim to show that the
use of recursion allows to construct very short and simple adaptive
quadrature code that is not inferior to such a sophisticated program as
QUANC8. Meanwhile the direct comparison of these programs seems to be
incorrect because of a number of reasons.

The Newton-Cotes equidistant quadrature formula which is used in \linebreak
QUANC8 allows to make reuse of integrand function values calculated in the
previous steps of bisection. That is the reason why QUANC8 has to have
higher performance in integrand function calls compared to adaptive
programs that use quadratures with non-equidistant points. Since QUADREC is
not oriented on the use of definite but any quadrature formula it can be
specified as a program of the later type.

At the same time QUANC8 gives bad results for functions with unlimitedly
growing derivative and does not work at all for functions that go to
infinity at the either of the integration interval endpoints. There are
none of the indicated restrictions in QUADREC. Furthermore the opportunity
of choosing of quadrature type makes it to be a very flexible tool for
integration. Here QUADREC gives a chance to choose quadrature which is the
most appropriate to the task (see Section \ref{optim}).

For integrals in sections \ref{section4_1} and \ref{section4_2} the optimal
numbers of quadrature points were found and used for integration. The
24-point quadrature was applied for integration in sections
\ref{section4_3} and \ref{section4_4}.

\subsection{Sharp peaks at a smooth background}
\label{section4_1}
Let us start with the calculation of the integral cited in \cite{Fors}:
\begin{equation}
\int\limits_0^1\left(\frac 1{(x-a_1)^2+a_2}+\frac 1{(x-b_1)^2+b_2}-
c_0\right)dx
\label{pr:2}\end{equation}

The integrand is the sum of two Lorenz type peaks and a constant
background. At the beginning, values of $a_1$, $a_2$, $b_1$, $b_2$ and
$c_0$ parameters were chosen to be the same as in the cited work. Then test
was conducted at decreasing values of $a_2$ and $b_2$, which determine
width of the peaks, while both programs satisfied desired accuracy and did
not signal about raw subintervals. The results of the test when
$a_2=b_2=10^{-8}$ are presented in Table \ref{tab:number2}. Note that only
the optimal values are given for QUADREC program. The corresponding optimal
numbers of Gaussian quadrature points are indicated.

\begin{table}[ht]
\centering
\caption{Testing results for integral (2).}\vspace{2mm}
\begin{tabular}{|c|c|c|c|c|c|} \hline
Desired   &\multicolumn{2}{|c|}{QUANC8}  &\multicolumn{3}{|c|}{QUADREC}
\\ \cline{2-6}
relative  & Number of     &  Reached &  Number of     & Reached &   Optimal\\
accuracy  & function calls&  accuracy&  function calls& accuracy&   quadrature
\\ \hline
 1.0e-4   & 433  & 5.8e-06  & 665  & 3.7e-08 &  7   \\
 1.0e-5   & 513  & 1.5e-06  & 721  & 3.8e-08 &  7   \\
 1.0e-6   & 641  & 5.0e-07  & 792  & 3.2e-09 &  8   \\
 1.0e-7   & 801  & 7.2e-08  & 952  & 4.6e-10 &  8   \\
 1.0e-8   & 993  & 2.8e-09  & 1080 & 1.6e-10 &  8   \\
 1.0e-9   & 1217 & 2.1e-10  & 1304 & 8.8e-13 &  8   \\
 1.0e-10  & 1521 & 4.6e-11  & 1399 & 2.4e-14 &  13  \\
 1.0e-11  & 1825 & 6.5e-12  & 1599 & 4.4e-15 &  13  \\
 1.0e-12  & 2337 & 1.3e-13  & 1807 & 2.9e-16 &  13  \\
 1.0e-13  & 2945 & 1.4e-14  & 1859 & 1.1e-18 &  13  \\
 1.0e-14  & 3665 & 8.1e-16  & 1911 & 4.3e-19 &  13  \\ \hline
\end{tabular}
\label{tab:number2}
\end{table}

As it follows from the given data the number of integrand function calls
are compared for the both programs in the wide range of desired accuracy.
Meanwhile it is interesting to point the attention on the fact that the use
of optimal quadratures can give definite profits in the reached accuracy
and number of integrand function calls even when the simple comparison rule
(\ref{pr:1}) and no reuse of the integrand values are applied.

At further decreasing of $a_2$ and $b_2$ parameters QUANC8 informed about
raw intervals and did not satisfy desired accuracy. At the same time
QUADREC gave correct results for the integral down to parameter values
$a_2=b_2=10^{-19}$. This is mainly due to the differences between static
and dynamic memory allocation ideologies which are used in QUANC8 and
QUADREC respectively.

\subsection{Integration at a given absolute accuracy}
\label{section4_2}
The next test concerns the integration of the function:
\begin{equation} f(x)={(1-\alpha x)\exp(-\alpha x)\over
x^2\exp(-2\alpha x)+b^2}
\label{pr:3}\end{equation}

over whole positive real axes. It is easy to show that its exact value is
equal to zero. For reducing the integration interval to [0,1] the evident
substitution x=t/(1-t) was used. As far as absolute accuracy of the
integration was required the fourteenth line in the listed above QUADREC
function text was changed to:

\vspace{3mm}
\verb| if (dabs(SumLeft+SumRight-Estimation).gt.Eps) then|
\vspace{3mm}

The results of the test are presented in Table \ref{tab:number3}.
\begin{table}[ht]
\centering
\caption{Testing results for integrand (3).}\vspace{2mm}
\begin{tabular}{|c|c|c|c|c|c|} \hline
Desired   &\multicolumn{2}{|c|}{QUANC8}  &\multicolumn{3}{|c|}{QUADREC}
\\ \cline{2-6}
absolute  & Number of     &  Reached &  Number of     & Reached &   Optimal\\
accuracy  & function calls&  accuracy&  function calls& accuracy&   quadrature
\\ \hline
10e-5   & 33  &  1.1e-06  &  44  & -3.7e-07   &  4   \\
10e-6   & 33  &  1.1e-06  &  55  & -7.1e-08   &  5   \\
10e-7   & 49  &  1.1e-07  &  75  & -1.1e-10   &  5   \\
10e-8   & 49  &  1.1e-07  &  95  & -3.5e-13   &  5   \\
10e-9   & 65  &  6.1e-11  & 114  & -1.9e-14   &  6   \\
10e-10  & 81  &  1.7e-13  & 114  & -1.9e-14   &  6   \\
10e-11  & 81  &  1.7e-13  & 133  & -2.2e-16   &  7   \\
10e-12  &145  & -5.6e-14  & 152  & -4.2e-18   &  8   \\
10e-13  &161  &  1.4e-15  & 152  & -4.2e-18   &  8   \\
10e-14  &193  &  6.7e-16  & 190  & -1.1e-19   & 10   \\
10e-15  &241  & -3.4e-17  & 209  & -1.2e-19   & 11   \\
10e-16  &305  & -3.0e-17  & 230  & -8.1e-20   & 10   \\
10e-17  &353  & -7.3e-19  & 253  & -1.2e-19   & 11   \\
10e-18  &513  &  4.1e-19  & 266  & -2.8e-20   & 14   \\ \hline
\end{tabular}
\label{tab:number3}
\end{table}

\subsection{Improper integral}
\label{section4_3}
The next integrand function:
\begin{equation} f(x)=x^{\frac 1n-1}  \label{pr:4}\end{equation}

becomes nonintegrable in the [0,1] interval when n goes to infinity.
Besides function \ref{pr:4} goes to infinity at the low integration limit.
That is the reason why QUANC8 can not be directly applied to the problem.
To have still the opportunity of comparison, the integration of indicated
function was performed from $10^{-10}$ to 1. The results of the test for
number n up to and including 20 and desired relative accuracy of $10^{-14}$
are listed in Table \ref{tab:number4}. The second and fifth columns give
the number of intervals that were not processed during the integration by
both routines. The number of integrand function calls and values of reached
relative accuracy are also presented in the table.

\begin{table}[h]
\centering
\caption{Testing results for integration of (4) over
[$10^{-10}$,1].}\vspace{2mm}
\begin{tabular}{|c|c|c|c|c|c|c|} \hline
   &\multicolumn{3}{|c|}{QUANC8}  &\multicolumn{3}{|c|}{QUADREC}      \\
\cline{2-7}
 n &   Raw     & Number of &  Reached &   Raw     &  Number of& Reached \\
   & intervals & function  &  accuracy& intervals &  function & accuracy \\
   &           &  calls    &          &           &  calls    &          \\
\hline
  1 &   0 &   33 & 0.0e+00 & 0 &   72 & 5.4e-20   \\
  2 &  14 & 2033 & 6.3e-10 & 0 & 2856 & 1.7e-18   \\
  3 &  22 & 2513 & 3.3e-08 & 0 & 2952 & 1.1e-16   \\
  4 &  50 & 3329 & 2.1e-07 & 0 & 2952 & 2.6e-18   \\
  5 &  70 & 3793 & 6.0e-07 & 0 & 2952 & 2.1e-16   \\
  6 &  74 & 3873 & 1.2e-06 & 0 & 2952 & 2.1e-16   \\
  7 & 108 & 4017 & 1.9e-06 & 0 & 2952 & 2.8e-16   \\
  8 & 130 & 4049 & 2.6e-06 & 0 & 2952 & 3.1e-18   \\
  9 & 150 & 4113 & 3.4e-06 & 0 & 2952 & 3.3e-16   \\
 10 & 146 & 4113 & 4.2e-06 & 0 & 2952 & 1.5e-16   \\
 11 & 164 & 4081 & 4.9e-06 & 0 & 3048 & 2.4e-16   \\
 12 & 176 & 4097 & 5.6e-06 & 0 & 3048 & 2.8e-16   \\
 13 & 164 & 4113 & 6.2e-06 & 0 & 3048 & 4.1e-16   \\
 14 & 184 & 4097 & 6.8e-06 & 0 & 3048 & 2.9e-16   \\
 15 & 200 & 4129 & 7.4e-06 & 0 & 3048 & 1.4e-16   \\
 16 & 166 & 4097 & 7.9e-06 & 0 & 3048 & 2.8e-18   \\
 17 & 187 & 4113 & 8.4e-06 & 0 & 3048 & 1.1e-16   \\
 18 & 182 & 4161 & 8.8e-06 & 0 & 3048 & 1.9e-16   \\
 19 & 184 & 4113 & 9.3e-06 & 0 & 3048 & 4.5e-16   \\
 20 & 156 & 4097 & 9.7e-06 & 0 & 3048 & 4.4e-16   \\ \hline
\end{tabular}
\label{tab:number4}
\end{table}

Such a low performance of QUANC8, as it follows from the given data, can be
explained by the type of comparison rule which it exploits. Really it is
not applicable to functions like \ref{pr:4} that have unlimitedly growing
derivative in some point(s) of integration interval.

The results of integration over whole interval [0,1] by QUADREC are shown
in Table \ref{tab:number5}. The last column contains the maximum recursion
level needed to achieve the desired accuracy. As far as it turned out to be
a rather big the program stack size was correspondingly increased.

\begin{table}[h]
\centering
\caption{Testing results for integration of (4) over
[0,1].}\vspace{2mm}
\begin{tabular}{|c|c|c|c|c|} \hline
 n &   Raw     & Number of        &  Reached  & Maximum         \\
   & intervals & function  calls  &  accuracy & recursion level \\ \hline
1  &  0  &    72 &  0.0e+00  &   1     \\
2  &  0  &  6504 &  1.0e-12  &  68     \\
3  &  0  & 10344 &  1.1e-12  & 108     \\
4  &  0  & 14088 &  1.2e-12  & 147     \\
5  &  0  & 17928 &  1.2e-12  & 187     \\
6  &  0  & 21672 &  1.3e-12  & 226     \\
7  &  0  & 25416 &  1.3e-12  & 265     \\
8  &  0  & 29256 &  1.3e-12  & 305     \\
9  &  0  & 33000 &  1.3e-12  & 344     \\
10 &  0  & 36744 &  1.4e-12  & 383     \\
11 &  0  & 40584 &  1.3e-12  & 423     \\
12 &  0  & 44328 &  1.4e-12  & 462     \\
13 &  0  & 48072 &  1.4e-12  & 501     \\
14 &  0  & 51912 &  1.4e-12  & 541     \\
15 &  0  & 55656 &  1.4e-12  & 580     \\
16 &  0  & 59400 &  1.4e-12  & 619     \\
17 &  0  & 63240 &  1.4e-12  & 659     \\
18 &  0  & 66984 &  1.4e-12  & 698     \\
19 &  0  & 70728 &  1.4e-12  & 737     \\
20 &  0  & 74568 &  1.4e-12  & 777     \\ \hline
\end{tabular}
\label{tab:number5}
\end{table}

\subsection{Evidence of program adaptation}
\label{section4_4}
Finally, integration of:
\begin{equation} f(x)=sin(Mx)
\label{pr:5}\end{equation}

over [0,2$\pi$] interval is appropriate for demonstrating the ability of an
integration routine for adaptation. The exact integral value evidently is
equal to zero for any integer M.

During the test a number of large simple integers were assigned to M and
absolute integration accuracy of $10^{-10}$ was required. The output of the
test is presented in Table  \ref{tab:number6}. The accommodation of the
routine evidently follows from the data. Namely, for M's going from
$\approx 10^5$ to $\approx 1.2\cdot 10^6$ the maximum recursion level
(integrand function call number) changes from 15 ($\approx 1.6\cdot 10^6$)
to 18 ($\approx  1.2\cdot 10^7$).

For all chosen values of M the desired accuracy was fulfilled. Furthermore
program succeeded down to the accuracy of $10^{-12}$. Note that the
standard stack size was used during the test and higher performance of the
QUADREC certainly could be reached had the stack size been enlarged.
\begin{table}[ht]
\centering
\caption{Testing results for integrand (5).}\vspace{2mm}
\begin{tabular}{|c|c|c|c|} \hline
 n & Maximum         & Number of        &  Reached absolute  \\
   & recursion level & integrand calls  &  accuracy  \\ \hline
100003  &   15 &   1572840 &   3.2e-14  \\
200003  &   16 &   3145704 &  -7.2e-14  \\
300007  &   16 &   3145704 &   7.0e-15  \\
400009  &   17 &   6290664 &   1.9e-14  \\
500009  &   17 &   6291432 &  -2.2e-14  \\
600011  &   17 &   6291432 &  -5.7e-15  \\
700001  &   17 &  12364392 &   1.3e-15  \\
800011  &   18 &  12580200 &   1.0e-13  \\
900001  &   18 &  12582888 &   9.7e-14  \\
1000003 &   18 &  12582888 &   2.2e-13  \\
1100009 &   18 &  12582888 &   2.4e-15  \\
1200007 &   18 &  12582888 &  -9.4e-14  \\ \hline
\end{tabular}
\label{tab:number6}
\end{table}

\section{The optimal quadrature}
\label{optim}
\begin{figure}[t]
\begin{center}
\psfig{figure=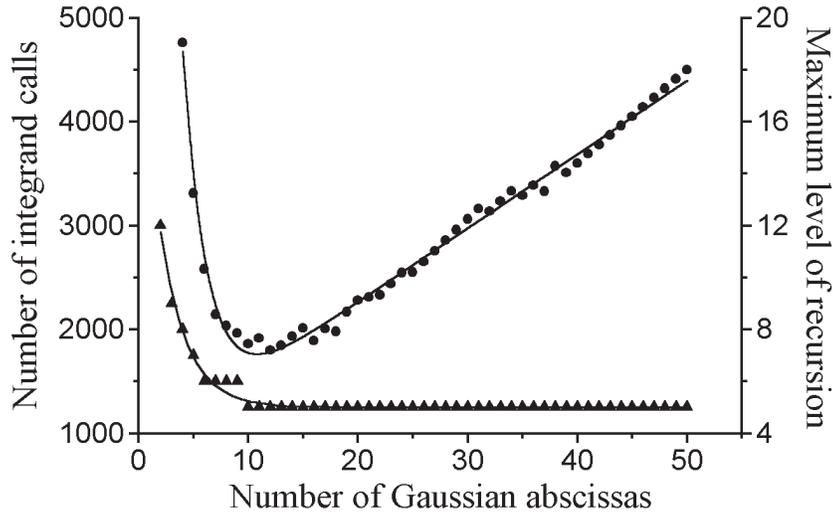,width=115mm}
\end{center}
\caption{\label{fig2}Number of integrand calls (circles) and maximum
recursion level (triangles) versus the number of Gaussian abscissas.}
\end{figure}

Here we want to take note of the possibility to minimize number of integrand
calls by choosing the quadrature with optimal number of abscissas. The
fig.\ref{fig2} demonstrates such a possibility.

\vspace{5mm}

Particularly, circles present the number of integrand calls versus the
number of Gaussian abscissas used for the integration of \ref{pr:2} over
[0,1] and desired relative accuracy of $10^{-10}$ and $a_1=0.3$,
$a_2=10^{-9}$, $b_1=0.9$, $b_2=4.0\cdot 10^{-9}$, $c_0=6$. From the
presented data one can see that the optimal number of Gaussian abscissas
ranges approximately from 7 to 17. The use of more Gaussian abscissas leads
to the linear growth of the number of function calls, because it does not
result in essential reduction of the integration interval fragmentation.
From the other hand the use of less number of Gaussian abscissas results in
the significant growth of the number of function calls due to the extremely
high fragmentation required. Dependence of the maximum recursion level upon
the number of abscissas, presented by triangles, confirms this
consideration.

Note that despite significant differences of the testing integrals
(\ref{pr:2},\ref{pr:3}) the optimal number of Gaussian abscissas turned out
to be in the above limits.

\section{Conclusion}
Thus, the indirect recursion combined with AUTOMATIC variable description
allow to employ true recursion mechanism in MS Fortran-77. In particular,
the recursion strategy was applied to create effective adaptive quadrature
code. Despite the simplicity and extremely small program body it showed
good results on rather complex testing integrals. The created subroutine is
very flexible and applicable to a wide range of integration problems. In
particular, it was applied for constructing effective Hilbert
transformation program. The last one was used to restore frequency
dependence of refraction coefficient in analysis of optical properties of
complex organic compounds. The subroutine can be easily incorporated into
existing Fortran programs. Note that the coding trick, described in the
paper, is very convenient for constructing multidimensional adaptive
quadrature programs.

\section{Acknowledgments}
We express our thanks to Dr.V.K.Basenko for stimulating and useful
discussion of the problem.

\newpage
\appendix{{\large\bf Appendix}}
\section{Guass-Legendr weights and quadratures}
\label{GL}
\begin{verbatim}
      subroutine gauleg(x1,x2,x,w,n)
      integer n
      real*8 x1,x2,x(n),w(n)
      real*8  eps
      parameter (eps=3.d-14)
      integer i,j,m
      real*8  p1,p2,p3,pp,xl,xm,z,z1
      m=(n+1)/2
      xm=0.5d0*(x2+x1)
      xl=0.5d0*(x2-x1)
      do 12 i=1,m
        z=cos(3.141592654d0*(i-.25d0)/(n+.5d0))
1       continue
          p1=1.d0
          p2=0.d0
          do 11 j=1,n
            p3=p2
            p2=p1
            p1=((2.d0*j-1.d0)*z*p2-(j-1.d0)*p3)/j
11        continue
          pp=n*(z*p1-p2)/(z*z-1.d0)
          z1=z
          z=z1-p1/pp
        if(abs(z-z1).gt.eps)goto 1
        x(i)=xm-xl*z
        x(n+1-i)=xm+xl*z
        w(i)=2.d0*xl/((1.d0-z*z)*pp*pp)
        w(n+1-i)=w(i)
12    continue
      return
      end
\end{verbatim}

\begin{thebibliography}{30}
\bibitem{Lynn} J.N.Lynnes, Comm. ACM 13(1970), p.260.
\bibitem{Genz} Genz and J.S.Chisholm, Computer PH 4(1972), p. 11 - 15.
\bibitem{Kah} Kahaner and M.B. Wells, SIAM Rev 18(1976), p. 811.
\bibitem{Fors} G.E.Forsythe,M.A.Malcolm,C.B.Moler, Computer methods for
mathematical computations  (Princeton Hall INC. 1977).
\bibitem{Lew} Lewellen, Computer Physics Communication 27(1982), p. 167 - 178.
\bibitem{Cor} Corliss and L.B Rall,  SIAM J SCI 8(1987), p.831 - 847.
\bibitem{Kron} Kronrod A.S.,1964. Doklady Akdemii Nauk SSSR, vol. 154,
p. 283-286.
\bibitem{Press} W.H.Press, S.A.Teukolsky, W.T.Vetterling, B.P.Flannery.
Numerical recipes in Fortran.
\bibitem{Mat} J.Mathews, R.L.Walker. Mathematical methods of physics.
W.A. Benjamin, INC. 1964.
\bibitem{BZh} A.N.Berlizov, A.A.Zhmudsky. The recursive one-dimensional
adaptive quadrature code. Preprint Institute for Nuclear Research. Kyiv. 1998.
\bibitem{Zh} A.A.Zhmudsky. One class of integrals evaluation in magnet
solitons theory. Preprint LANL. /9800312. 1998.
\end{thebibliography}
\end{document}